\documentclass[12pt,thmsa]{article}
\usepackage{amsmath}
\usepackage{amssymb}
\usepackage{amsfonts}
\usepackage[T1]{fontenc}

\begin{document}

\title{On the Physical Meaning of the Robinson-Trautman-Maxwell Fields}
\author{Carlos Kozameh$^{1}$ \and E.T. Newman$^{2}$ \and Gilberto
Silva-Ortigoza$^{3} $ \\
$^{1}$FaMaF, Univ. of Cordoba, \\
Cordoba, Argentina\\
$^{2}$Dept of Physics and Astronomy, \\
Univ. of Pittsburgh, \\
Pittsburgh, PA 15260, USA\\
$^{3}$Facultad de Ciencias F\'{\i}sico Matem\'{a}ticas \\
de la Universidad Aut\'{o}noma de Puebla, \\
Apartado Postal 1152, 72001,\\
Puebla, Pue., M\'{e}xico}
\date{7.12.06 }
\maketitle

\begin{abstract}
We study the Robinson-Trautman-Maxwell Fields in two closely related
coordinate systems, the original Robinson-Trautman (RT) coordinates (in a
more general context, often referred to as NU coordinates) and Bondi
coordinates. In particular, we identify one of the RT variables as a
velocity and then from the Bondi energy-momentum 4-vector, we find kinematic
expressions for the mass and momentum in terms of this velocity. From these
kinematic expressions and the energy-momentum loss equation we obtain
surprising equations of motion for `the center of mass' of the source where
the motion takes place in the four-dimensional Poincare translation
sub-group of the BMS group.
\end{abstract}

\section{\protect\bigskip}

Introduction

One of the more important discoveries concerning integrating the Einstein
equations was the realization that the condition of metrics being
algebraically special allowed a huge simplification in the integration\cite%
{RT}. Not only was there the beautiful geometric result, the Goldberg-Sachs
Theorem\cite{G.S.}, saying that the degenerate principal null vectors must
form a null geodesic congruence that is shear-free, but this result allowed
all radial integrations to be performed. The algebraically special metrics
separate into two classes, those with the null geodesic congruence having a
vanishing\cite{RT} twist and those with non-vanishing\cite{Talbot,Lind,Tod}
twist. We will study the subclass of the twist-free congruences, with a
double degeneracy, known as the type II Robinson-Trautman (RT) metrics. Many
of the issues explored here have been studied earlier by other authors\cite%
{RT,Posadas,Lind,Tod,Drey,Tafel}. In particular Tafel and Drey looked at the
relationship between the RT metrics and the Bondi metrics.

It is the purpose of this note to describe some rather surprising results
concerning the physical content that is contained in both the type II vacuum
Robinson-Trautman\cite{RT} and Robinson-Trautman-Maxwell\cite{Lind,Tod}
metrics, but had been essentially unobserved. Since, at least formally, the
RT spaces are special cases of the RTM spaces, often we will just refer to
both as RTM though there are apparently significant subtle issues that
separate them.

These metrics are \textit{largely} determined by a single function of three
variables denoted by $V=V(\tau ,\zeta ,\overline{\zeta }),$ where $\tau $ is
the NU/RT time\cite{Tod,Gil.Ted} coordinate and ($\zeta ,\overline{\zeta }$)
are complex stereographic coordinates on the sphere. (The coordinates used
by Robinson-Trautman are a special case of those referred to as NU
coordinates.) $V(\tau ,\zeta ,\overline{\zeta }),$ whose geometric meaning
will be clarified later, satisfies the Robinson-Trautman equation or its
extension to include the Maxwell field.

The basic plan in this work is to study the consequences of expanding $V$ in
spherical harmonics, i.e., assume that
\begin{equation}
V(\tau ,\zeta ,\overline{\zeta })=\sum V^{lm}(\tau )Y_{lm}(\zeta ,\overline{
\zeta }),  \label{0}
\end{equation}
and using the expansion in the RTM equations. The fact that the RTM
equations are very non-linear means that the spherical harmonic
decomposition of the equation itself is quite complicated and that
approximations were necessary. The approximated RTM equations for $V$ then
become a system of coupled ODE's for the evolution of the components $%
V^{lm}. $ The higher spherical harmonic components, i.e., those for $%
l\eqslantgtr 2,$ become derivatives of multiple moments. Our interest
(partially) centers on the lowest components, i.e., $(l=0,1),$ which can be
interpreted as the \textit{four-velocity of a world-line} in the
four-dimensional space of the Poincar\'{e} translation subgroup of the BMS
group. The equations of motion for this world-line are contained in the
lowest harmonic part of the RTM equations but are driven by the higher
harmonics$.$

The surprise comes when these equations are translated into Bondi
coordinates and one looks at the Bondi energy-momentum four-vector and its
evolution; namely the energy-momentum loss equation. Though the
transformation to the Bondi coordinates is complicated and can be done in
practice only perturbatively in powers of $v/c$, nevertheless we see a
beautiful confirmation of the identification of the $l=1$ components with
the three-velocity $v^{i}$. The Bondi three-momentum becomes, in the first
order approximation
\begin{equation}
P^{i}=M_{0}v^{i}-\frac{2}{3c^{3}}q^{2}v^{i\,\prime }.  \label{1}
\end{equation}
In other words we obtain the usual kinematic definition of the momentum but
now (by including the Maxwell field) with the standard radiation reaction
term. In addition the Bondi energy, in the second order approximation,
contains the relativistic kinematic mass-energy relation
\begin{equation}
E=Mc^{2}=M_{0}c^{2}(1+\frac{1}{2}\frac{v^{2}}{c^{2}})\simeq \frac{M_{0}c^{2}%
}{\sqrt{1-\frac{v^{2}}{c^{2}}}}.  \label{2}
\end{equation}

Higher moment contributions have been left out in both $E$ and $P^{i}.$

The calculations that were involved in order to obtain these results were
fairly lengthy and complicated and far from obvious. Since the equations are
quite non-linear, considerable use had to be made of Clebsch-Gordon
expansions of spherical harmonic products. In order to keep the presentation
as clear as possible we have the following organization. In the second
section we review our notation. In section three we derive the \textit{\
flat-space metric} based on the null coordinates that are associated with
the light-cones with apex on an \textit{arbitrary} time-like world-line in
Minkowski space. The variable $V(\tau ,\zeta ,\overline{\zeta })$ appears
naturally in this metric and one sees in this simple case how the velocity
associated with the world-line enters the $V(\tau ,\zeta ,\overline{\zeta })$
via the $l=(0,1)$ harmonics. In this flat space case the coefficients of the
$l=(0,1)$ are essentially arbitrary while in our RT and RTM cases these
coefficients satisfy equations of motion and are driven by the higher
harmonics. In section four we state the RTM metrics and their related
differential equations and discuss their origin and how they are related to
the general asymptotically flat Einstein-Maxwell metrics. In section five we
state the details of our approximations and show how in this context the RT
and RTM equations can be integrated. In section six we describe the
transformation from the NU/RT coordinates to Bondi coordinates and construct
the Bondi energy-momentum four vector and its evolution equation. Finally in
the concluding section we discuss how this work fits into a much broader -
e.g., to the twisting algebraically special type II metrics and even to a
large class of algebraically general asymptotically flat Einstein-Maxwell
fields.

The calculations described here were often long and tedious and short-cuts
were not found. This, unfortunately, does make the work difficult to follow
in detail. To try to compensate for this we have attempted to give
considerable background material.

\section{Notation}

Throughout this work we will be using certain Lorentzian and Euclidean
vectors and tensor valued functions on the sphere. The most important of
them is the null vector $l^{a}=\eta ^{ab}l_{b}$ that sweeps out the entire
sphere of null directions as
\begin{equation}
(\zeta ,\overline{\zeta })=cot\frac{\theta }{2}(e^{i\phi },e^{-i\phi })
\label{stereo}
\end{equation}
ranges over the sphere. Our space-time conventions are such that
\begin{equation*}
\eta ^{ab}=\eta _{ab}=diag(1,-1,-1,-1).
\end{equation*}
In a given Lorentz frame we chose $l^{a}$ to have the normalized form
\begin{equation}
l^{a}=\frac{\sqrt{2}}{2}(1,\frac{\zeta +\overline{\zeta }}{1+\zeta \overline{
\zeta }},-i\frac{\zeta -\overline{\zeta }}{1+\zeta \overline{\zeta }},\frac{
-1+\zeta \overline{\zeta }}{1+\zeta \overline{\zeta }})=\frac{\sqrt{2}}{2}
(t^{a}+C^{a})  \label{l^}
\end{equation}
$t^{a}$ and $C^{a}$ being unit time and space-like vectors with $%
t^{a}=(1,0,0,0).$ The factor $\frac{\sqrt{2}}{2}$ had been chosen
historically so that the antipodal null vector $n^{a},$ with $n^{a}l_{a}=1,$
is `symmetrically' defined, i.e.,
\begin{equation}
n^{a}=\frac{\sqrt{2}}{2}(1,-\frac{\zeta +\overline{\zeta }}{1+\zeta
\overline{\zeta }},i\frac{\zeta -\overline{\zeta }}{1+\zeta \overline{\zeta }%
},\frac{1-\zeta \overline{\zeta }}{1+\zeta \overline{\zeta }})=\frac{\sqrt{2}%
}{2}(t^{a}-C^{a}).  \label{n^}
\end{equation}

{\bf Remark 1.} {\it This normalization, which has been in use for
many years and has many formal benefits, nevertheless has certain
serious drawbacks, namely the appearance of some unpleasant factors
of 2 and $\sqrt{2}.$ Unfortunately they will be often seen and will
create a serious annoyance. They appear, mainly, in the definitions
of the several time coordinates.}\\

Note that for the covariant forms of $l^{a}$ and $n^{a},$ we just change the
signs in the last three components, e.g.,
\begin{equation}
l_{a}=\frac{\sqrt{2}}{2}(1,-\frac{\zeta +\overline{\zeta }}{1+\zeta
\overline{\zeta }},i\frac{\zeta -\overline{\zeta }}{1+\zeta \overline{\zeta }%
},\frac{1-\zeta \overline{\zeta }}{1+\zeta \overline{\zeta }}).  \label{l_}
\end{equation}

Other (co)vectors obtained directly from the $l_{a}$ are
\begin{eqnarray}
m_{a} &=&\eth l_{a}=\frac{\sqrt{2}}{2}(0,-\frac{1-\overline{\zeta }^{2}}{%
1+\zeta \overline{\zeta }},\frac{i(1+\overline{\zeta }^{2})}{1+\zeta
\overline{\zeta }},\frac{-2\overline{\zeta }}{1+\zeta \overline{\zeta }}),
\label{m} \\
\overline{m}_{a} &=&\overline{\eth }l_{a}=\frac{\sqrt{2}}{2}(0,-\frac{%
1-\zeta ^{2}}{1+\zeta \overline{\zeta }},-\frac{i(1+\zeta ^{2})}{1+\zeta
\overline{\zeta }},\frac{-2\zeta }{1+\zeta \overline{\zeta }}),  \label{mbar}
\\
t^{a} &=&\frac{\sqrt{2}}{2}(l^{a}+n^{a})=(1,0,0,0)=t_{a},  \label{t} \\
c_{a} &=&l_{a}-n_{a}=\sqrt{2}(0,-\frac{\zeta +\overline{\zeta }}{1+\zeta
\overline{\zeta }},i\frac{\zeta -\overline{\zeta }}{1+\zeta \overline{\zeta }%
},\frac{1-\zeta \overline{\zeta }}{1+\zeta \overline{\zeta }})=\sqrt{2}C_{a},
\label{c}
\end{eqnarray}

so that we can write
\begin{equation}
l_{a}=\frac{\sqrt{2}}{2}(t_{a}+\frac{c_{a}}{\sqrt{2}})=\frac{1}{\sqrt{2}}%
t_{a}+\frac{1}{2}c_{a}.  \label{t+c}
\end{equation}

We have that

\begin{equation*}
\eth \eta _{(s)}\equiv P_{0}^{1-s}\partial _{\zeta }(P_{0}^{s}\eta
_{(s)}),\qquad \overline{\eth }\eta _{(s)}\equiv P_{0}^{1+s}\partial _{%
\overline{\zeta }}(P_{0}^{-s}\eta _{(s)}),\qquad P_{0}=1+\zeta \overline{%
\zeta }.
\end{equation*}

Note that in the given Lorentzian frame, we can treat $m_{a},$ $\overline{m}%
_{a},$ and $c_{a}$ as Euclidean three-vectors which we write, with Euclidean
indices $i,j.....$ , as
\begin{equation*}
m_{i},\text{ }\overline{m}_{i},\text{ and }c_{i}.
\end{equation*}
Also note that from the definitions of the stereographic coordinates, Eq.(%
\ref{stereo}), we have that
\begin{equation*}
c_{i}=l_{i}-n_{i}=-\sqrt{2}(\cos \phi \sin \theta ,\sin \phi \sin \theta
,\cos \theta )=-\sqrt{2}C^{i}
\end{equation*}
where $C^{i}$ is the unit Euclidean radial vector.

The unifying notation of the tensor spin-$s$ harmonics\cite{spins}, $%
Y_{l\,ij....k.}^{s}$ $\Leftrightarrow $ $_{s}Y_{l\,m.....}$, which allows us
to write
\begin{eqnarray}
Y_{0}^{0} &=&1,  \label{00} \\
Y_{1i}^{1} &=&m_{i},  \label{11} \\
Y_{1i}^{0} &=&\overline{\eth }Y_{1i}^{1}=\eth Y_{1i}^{-1}=-c_{i},  \label{10}
\\
Y_{1i}^{-1} &=&\overline{m}_{i},  \label{-11}
\end{eqnarray}

and
\begin{eqnarray}
Y_{2ij}^{2} &=&m_{i}m_{j},  \label{22} \\
Y_{2ij}^{1} &=&\overline{\eth }Y_{2ij}^{2}=-(c_{i}m_{j}+m_{i}c_{j}),
\label{21} \\
Y_{2ij}^{0} &=&\overline{\eth }Y_{2ij}^{1}=3c_{i}c_{j}-2\delta _{ij},
\label{20} \\
Y_{2ij}^{-1} &=&-(c_{i}\overline{m}_{j}+\overline{m}_{i}c_{j}),  \label{2-1}
\\
Y_{2ij}^{-2} &=&\overline{m}_{i}\overline{m}_{j},  \label{2-2}
\end{eqnarray}%
turns out to be almost indispensable. The ordinary spherical harmonics $%
Y_{lm}$ are equivalent, i.e., are linear combinations of the $s=0$, $%
Y_{l\,ij..k}^{0},$%
\begin{equation*}
Y_{lm}\Leftrightarrow Y_{l\,ij..k}^{0}
\end{equation*}%
so that, for example
\begin{equation}
Y_{1m}=(Y_{11,}Y_{10,}Y_{1-1})\Leftrightarrow Y_{1\,i}^{0}=-c_{i}.
\label{Y_1}
\end{equation}

Due to the non-linearity of the RTM equations, products of these harmonics
often appear and must be reduced via a Clebsch-Gordon expansion. Several of
the most important and most used of these expansions\cite{spins} are:

\begin{eqnarray}
Y_{1i}^{1}Y_{1j}^{-1} &=&\frac{1}{3}\delta _{ij}-\frac{i\sqrt{2}}{4}\epsilon
_{ijk}Y_{1k}^{0}-\frac{1}{12}Y_{2ij}^{0},  \label{YY1} \\
Y_{1i}^{0}Y_{1j}^{0} &=&\frac{2}{3}\delta _{ij}+\frac{1}{3}Y_{2ij}^{0},
\label{YY2}
\end{eqnarray}
and

\begin{eqnarray}
Y_{2kl}^{2}Y_{2ij}^{-2} &=&\frac{1}{5}\{\frac{1}{2}\delta _{ik}\delta _{jl}+%
\frac{1}{2}\delta _{li}\delta _{kj}-\frac{1}{3}\delta _{ij}\delta _{kl}\}
\label{YY3} \\
&&+\frac{2}{5}\{i\frac{\sqrt{2}}{8}(\delta _{jl}\epsilon _{ike}+\delta
_{ki}\epsilon _{jle}+\epsilon _{jke}\delta _{il}+\delta _{kj}\epsilon
_{ile})Y_{1e}^{0}\}  \notag \\
&&+\frac{2}{7}\{\frac{1}{6}(\delta _{ij}Y_{2kl}^{0}+\delta _{kl}Y_{2ij}^{0})-%
\frac{1}{8}(\delta _{lj}Y_{2ik}^{0}+\delta _{ki}Y_{2lj}^{0}+\delta
_{li}Y_{2kj}^{0}+\delta _{kj}Y_{2il}^{0})\}  \notag \\
\text{+ (}l\text{{}} &=&\text{3 and 4 harmonics).}  \notag
\end{eqnarray}

Using the tensor spin-$s$ harmonics, the spin-$s$ functions can be expanded
via the $Y_{l\,ij..k}^{s}$. In particular, for the ordinary functions on the
sphere, i.e., for $s=0$ functions, we have that
\begin{equation*}
F_{(0)}=f^{0}Y_{0}^{0}+f^{i}Y_{1i}^{0}+f^{ij}Y_{2ij}^{0}+...,
\end{equation*}
rather than the equivalent expansion in terms of the ordinary spherical
harmonics. A particular function that often appears is
\begin{equation}
G_{(0)}=g^{a}l_{a}=\frac{1}{\sqrt{2}}g^{0}-\frac{1}{2}g^{i}Y_{1i}^{0},
\label{example1}
\end{equation}
where we have used Eqs.(\ref{t+c}) and (\ref{10}).

We will be using several different time variables (or coordinates) in the
course of this work, i.e., $\tau ,$ $w,$ $u_{B}$. The variable $\tau $ is
the proper time along a four-dimensional world-line; its derivative is
denoted by $\partial _{\tau }=(^{\prime }).$ Often we will be dealing with
functions of $w$ and of $\tau $ that functionally are identical, i.e., $%
F(\tau )$ and $F(w)$ are identical functions. In that case we will again use
the same notation, $\partial _{w}=(^{\prime })$ with the meaning coming from
the context. The (rescaled) Bondi time, $u_{B}$ $=w/\sqrt{2}$ will be
denoted by $\partial _{u_{B}}=(^{\cdot }).$ The origin of the$\sqrt{2}$ is
related to remark 1.

\section{A Flat Metric}

Before considering the RTM metrics we describe a flat-space version that
imitates many of the features of the RTM metrics, albeit with a great
simplification. Starting with Minkowski space and ordinary Minkowski
coordinates, $x^{a},$ we introduce a coordinate transformation to new
coordinates ($\tau ,r,\zeta ,\overline{\zeta }$) that are based on the null
cones emanating from an arbitrary time-like world line which \textit{we
parametrize so that it has a unit velocity vector,}
\begin{equation}
v^{a}{}v_{a}=1,  \label{unit}
\end{equation}
so that (with our approximation. i.e., for slow motion)
\begin{eqnarray}
v^{0} &=&\frac{1}{\sqrt{(1-\frac{v^{2}}{c^{2}})}}\approx 1+\frac{1}{2}\frac{
v^{i\,2}}{c^{2}}=1+\delta v^{0},  \label{v^0} \\
\xi ^{0}(\tau ) &=&\int v^{0}d\tau =\tau +\delta \xi ^{0}.  \label{x^0}
\end{eqnarray}

The transformation is given by
\begin{equation}
x^{a}=\xi ^{a}(\tau )+\frac{r}{V_{0}}l^{a}(\zeta ,\overline{\zeta }),
\label{coor.trans}
\end{equation}
with $x^{a}=\xi ^{a}(\tau ),$ the parametrized world-line, $l^{a}(\zeta ,%
\overline{\zeta }),$ the null vector of Eq.(\ref{l^}) and $V_{0}$($\tau
,\zeta ,\overline{\zeta }$) defined as the $\tau $ dependent spin-weight $s$
$=0$ function on the sphere
\begin{eqnarray}
V_{0} &\equiv &\xi ^{a}{}^{\prime }l_{a}=v^{a}{}l_{a},  \label{V} \\
\xi ^{a}{}^{\prime } &=&\frac{d\xi ^{a}}{d\tau }.  \notag
\end{eqnarray}

{\bf Remark 2.} {\it
If the world-line was simply the time axis, i.e., $\xi ^{a}(u_{B})$ $%
=u_{B}t^{a},$ then the transformation, $x^{a}\Leftrightarrow $ $(u_{B},%
\widehat{r},\widehat{\zeta },\widehat{\overline{\zeta }})$ is given by
\begin{equation}
x^{a}=u_{B}t^{a}+\sqrt{2}\widehat{r}l^{a}(\widehat{\zeta },\widehat{%
\overline{\zeta }}).  \label{Bondi0}
\end{equation}
In this special case, the new coordinates are the (flat-space) Bondi
coordinates. }

To construct the new form of the metric we first take the differentials of
Eq.(\ref{coor.trans}) and simplify it:

\begin{equation*}
dx^{a}=\xi ^{a}{}^{\prime }d\tau +d(\frac{r}{V_{0}})l^{a}(\zeta ,\overline{
\zeta })+\frac{r}{V_{0}}dl^{a}(\zeta ,\overline{\zeta }).
\end{equation*}
Using
\begin{eqnarray*}
dl^{a} &\equiv &P_{0}^{-1}(m^{a}d\zeta +\overline{m}^{a}d\overline{\zeta }),
\\
dV_{0} &=&V_{0}^{\prime }d\tau +P_{0}^{-1}v^{a}(m_{a}d\zeta +\overline{m}%
_{a}d\overline{\zeta }),
\end{eqnarray*}
from Eqs.(\ref{m}) and (\ref{mbar}), we have that

\begin{eqnarray}
dx^{a} &=&v^{a}{}d\tau +l^{a}(\frac{dr}{V_{0}}-\frac{V_{0}^{\prime }}{%
V_{0}^{2}}rd\tau )  \label{dx} \\
&&-\frac{r}{V_{0}^{2}P_{0}}l^{a}v^{b}(m_{b}d\zeta +\overline{m}_{b}d%
\overline{\zeta })+\frac{r}{V_{0}P_{0}}(m^{a}d\zeta +\overline{m}^{a}d%
\overline{\zeta }),  \notag
\end{eqnarray}%
which is then substituted into $ds^{2}=\eta _{ab}dx^{a}dx^{b}.$ Using the
scalar products $v^{a}l_{a}=V_{0}$, $m^{a}\overline{m}_{a}=1$ and $%
m^{a}l_{a}=0$ we obtain [after the cancelation of two terms both containing $%
v^{b}(m_{b}d\zeta +\overline{m}_{b}d\overline{\zeta })$] the flat metric
\begin{equation}
ds^{2}=\eta _{ab}dx^{a}dx^{b}=(1-2\frac{V_{0}^{\prime }}{V_{0}}r)d\tau
^{2}+2d\tau dr-r^{2}\frac{2d\zeta d\overline{\zeta }}{V_{0}^{2}P_{0}^{2}},
\label{flat metric}
\end{equation}%
or (with $P=\frac{P_{0}V_{0}}{\sqrt{2}}$)
\begin{equation}
ds^{2}=\eta _{ab}dx^{a}dx^{b}=(1-2\frac{V_{0}^{\prime }}{V_{0}}r)d\tau
^{2}+2d\tau dr-r^{2}\frac{d\zeta d\overline{\zeta }}{P^{2}}.
\label{flat metric2}
\end{equation}%
The Gaussian curvature of the $r=constant$ surfaces is, after a brief
calculation,
\begin{equation*}
K=4P^{2}\partial _{\zeta }\partial _{\overline{\zeta }}\log P=\eth \overline{%
\eth }\text{log}P_{0}=1,
\end{equation*}%
the curvature of the unit sphere metric.

{\bf Remark 3.} {\it It should be emphasized that in the present
flat-space case we have the expressions:
\begin{eqnarray}
X_{0} &=&\xi ^{a}(\tau ){}l_{a}=\frac{1}{\sqrt{2}}(\tau +\delta \xi ^{0})-%
\frac{1}{2}\xi ^{i}Y_{1i}^{0},  \label{X1} \\
V_{0} &=&v^{a}(\tau ){}l_{a}=\frac{1}{\sqrt{2}}(1+\delta v^{0})-\frac{1}{2}%
v^{i}Y_{1i}^{0},  \label{V1} \\
V_{0}^{\prime } &=&v^{a\prime }(\tau ){}l_{a}=\frac{1}{\sqrt{2}}\delta
v^{0\prime }-\frac{1}{2}v^{i\prime }Y_{1i}^{0}\text{.}  \label{V'1}
\end{eqnarray}

They will be generalized to include higher harmonics in the next
section. The metric form will appear quite similar. }

\quad To make this metric, (\ref{flat metric2}), agree with a certain
conventional form\cite{Tod} (coming from the $l^{a}$ normalization) that has
been frequently used, we make the following simple rescaling of the $r$ and $%
\tau $

\begin{eqnarray}
r &=&\frac{r^{*}}{\sqrt{2}}\text{ }\&\text{ }\tau =\sqrt{2}\tau ^{*},
\label{rescale1} \\
v^{a} &=&\frac{v^{*\,a}}{\sqrt{2}},  \notag \\
V_{0} &=&\frac{V_{0}^{*}}{\sqrt{2}},  \notag
\end{eqnarray}
resulting in
\begin{eqnarray}
ds^{2} &=&2(K-\frac{V_{0}^{*\prime }}{V_{0}^{*}}r^{*})d\tau ^{*2}+2d\tau
^{*}dr^{*}-\frac{r^{*\,\,2}}{2}\frac{d\zeta d\overline{\zeta }}{P^{2}}\text{
,}  \label{flat metric3} \\
P &=&\frac{P_{0}V_{0}}{\sqrt{2}}=\frac{P_{0}V_{0}^{*}}{2}, \\
K &=&4P^{2}\partial _{\zeta }\partial _{\overline{\zeta }}\log P=1,
\end{eqnarray}
with
\begin{equation*}
V_{0}^{*\prime }\text{{}}=\text{{}}\partial _{\tau ^{*}}V_{0}^{*}.
\end{equation*}

Note that we continue to use ($^{\prime }$) for both $\tau $ and $\tau ^{*}$
derivatives. The distinction should be clear from the context.

As we remarked earlier for the flat case, our function $V_{0}=v^{a}(\tau
){}l_{a}$ will be generalized in the RTM case to include higher harmonics.
In the flat case, $V$ codes the information of the velocity vector of the
Minkowski space world-line. In the RTM case there is no such `real
world-line' but nevertheless a `world-line' in an auxiliary space does
exist. It is easiest to show its meaning first in the flat space case. To do
so we first return to the flat space Bondi coordinates $(u_{B},\widehat{r},%
\widehat{\zeta },\widehat{\overline{\zeta }}),.$ Eq.(\ref{Bondi0}).
Rescaling $\widehat{r}$ and $u_{B}$ as in Eq.(\ref{rescale1})

\begin{equation}
x^{a}=\sqrt{2}u_{B}^{*}t^{a}+\widehat{r}^{*}l^{a}(\widehat{\zeta },\widehat{%
\overline{\zeta }}).  \label{Bondi}
\end{equation}
The relationship between $(u_{B}^{*},\widehat{r}^{*},\widehat{\zeta },%
\widehat{\overline{\zeta }})$ and $(\tau ,r,\zeta ,\overline{\zeta })$ is
given by
\begin{equation}
\sqrt{2}u_{B}^{*}t^{a}+\widehat{r}^{*}l^{a}(\widehat{\zeta },\widehat{%
\overline{\zeta }})=\xi ^{a}(\tau )+\frac{r}{V_{0}}l^{a}(\zeta ,\overline{%
\zeta }),  \label{Bondi2*}
\end{equation}
which with a bit of effort can actually be disentangled as a coordinate
transformation. We need it only at future null infinity, $\mathfrak{I}^{+}$,
i.e., in the limit as $r\rightarrow \infty ,$ where $(\widehat{\zeta },%
\widehat{\overline{\zeta }})\rightarrow (\zeta ,\overline{\zeta }).$

In this limit, if (\ref{Bondi2*}) is multiplied by $l_{a}$ we obtain the
relationship
\begin{equation}
u_{B}^{*}=\xi ^{a}(\tau )l_{a}(\zeta ,\overline{\zeta }).  \label{u<tau}
\end{equation}
But since the 4-$d$ Poincar\'{e} translation subgroup of the BMS group is
given by
\begin{equation*}
u_{B}^{*}=d^{a}l_{a}(\zeta ,\overline{\zeta }),
\end{equation*}
we see immediately that we can interpret the `world-line' as a curve in the
space of BMS translations, $d^{a}$.

We will see how this type of $`$world-line' arises naturally in the study of
the RTM spaces.

{\bf Remark 4.} {\it Often there are two different time-coordinates
that are used at null infinity, $\mathfrak{I}^{+}\mathfrak{,}$
namely$\mathfrak{,}$ the Bondi time $u_{B}$ and the so-called NU
time $\tau .$ Roughly speaking, the 2-surfaces of constant $u_{B}$
are uniformly stacked above each other while the constant $\tau $
surfaces have an arbitrary uneven stacking. They are
related by an expression of the form, $u_{B}=X(\tau ,\zeta ,\overline{\zeta }%
).$ Eq.(\ref{u<tau}) is an example. This description applies to both
flat and asymptotically flat space-times. }

\section{RT and RTMaxwell Space-Times}

\subsection{The metric and differential equations}

The RTMaxwell metric, \textit{after all the radial integrations have been
performed}, can be put into the form\cite{RT,Lind,Tod}, (see Eq.(\ref{flat
metric3}))

\begin{equation}
ds^{2}=2(K-\frac{V^{*\prime }}{V^{*}}r^{*}+\frac{V^{*3}\psi _{2}^{*0}}{r^{*}}
+\frac{V^{*4}\phi _{1}^{*\,\,0}\overline{\phi }_{1}^{*\,\,0}}{r^{*2}})d\tau
^{*\,2}+2d\tau ^{*}dr^{*}-\frac{r^{*2}}{2}\frac{d\zeta d\overline{\zeta }}{
P^{2}},  \label{RTMmetric}
\end{equation}
with

\begin{eqnarray}
K &=&4P^{2}\partial _{\overline{\zeta }}\partial _{\zeta }\log P,  \label{K}
\\
P &=&\frac{V^{*}P_{0}}{2},  \label{P} \\
\psi _{2}^{*0} &=&-V^{*-3}\chi ,  \label{psi2}
\end{eqnarray}
and Maxwell field

\begin{eqnarray}
\phi _{0}^{*} &=&0,  \notag \\
\phi _{1}^{*} &=&\phi _{1}^{*0}r^{-2}\equiv \frac{q}{2}V^{*-2}r^{-2},
\label{coulomb} \\
\phi _{2}^{\,\,*} &=&\phi _{2}^{*\,\,0}r^{-1}.  \notag
\end{eqnarray}

The two gravitational variables, $V^{*}$ and $\chi \equiv -V^{*3}\psi
_{2}^{*0},$ and the Maxwell field, $\phi _{2}^{*\,\,0}$, are functions only
of ($\tau ^{*},\zeta ,\overline{\zeta }$), and satisfy the three coupled
differential equations, the RTM equations,

\begin{eqnarray}
\phi _{1}^{\ast 0\prime }+\eth _{(\tau ^{\ast })}[V^{\ast }\phi _{2}^{\ast 0}] &=&0,  \label{max} \\
\eth _{(\tau ^{\ast })}\chi +2kV^{\ast 3}\phi _{1}^{\ast 0}\overline{\phi }_{2}^{\ast 0} &=&0,  \label{RT} \\
(V^{\ast -3}\chi )^{\prime }-\{\eth _{(\tau ^{\ast })}^{2}
\overline{\eth }_{(\tau ^{\ast })}^{2}V^{\ast }-V^{\ast -1}
\overline{\eth }_{(\tau ^{\ast })}^{2}V^{\ast }\cdot \eth _{(\tau
^{\ast })}^{2}V^{\ast }\}+kV^{\ast }\phi _{2}^{\ast 0}
\overline{\phi}_{2}^{\ast 0} &=&0.  \label{RT2}
\end{eqnarray}%
The constant $q,$ in Eq.(\ref{coulomb}), is the conserved electric charge
while $\chi \ ($or $\psi _{2}^{\ast 0})$ carries the information of the
Bondi energy-momentum four-vector which will be extracted later, $%
k=2G/c^{4}, $ and \dh $_{(\tau ^{\ast })}$ is the edth operator holding $%
\tau ^{\ast }$ constant. The quantities $\phi _{1}^{\ast 0}$ and $\phi
_{2}^{\ast 0}$ are asymptotic tetrad components of the Maxwell field while $%
\chi \ ($or $\psi _{2}^{\ast 0})$ are asymptotic tetrad components of the
Weyl tensor.

It is useful to note the logic of these equations: They can be solved (in
principle) individually, in order. The first allows $\phi _{2}^{*0}$ to be
expressed as a function of $V$ while the second does the same for $\chi .$
Finally, the last determines the $\tau $-evolution of $V.$

{\bf Remark 5.} {\it The RTM equations are extremely difficult
equations to solve or analyze rigorously. Very few exact solutions
have been found. In the special case when $q=0$, i.e., the pure RT
equations, Piotr Chrusciel\cite{Piotr} showed that, with
sufficiently `nice' initial data, the solutions evolve
asymptotically, in time, to the Schwarzschild metric. In the general
case, though one would hope for a theorem saying the evolution would
lead to the Reissner-Nordstrom solution, this apparently has not yet
been shown. It does make an attractive conjecture. Later, in our
discussion of the approximate solutions, we can see how this issue
arises. }

{\bf Remark 6.} {\it For the time being, to streamline the notation
and avoid a plethora of similar symbols, we will drop the ($^{*})$
on the $V$ and $\tau .$ Later, to see the physical relevance of our
results, we must remember to restore the original $V$ and $\tau $
via Eq.(\ref{rescale1}). In addition, the velocity
of light, `$c$', will be explicitly introduced via $\tau \Rightarrow $ $%
c\tau $ and $w\Rightarrow cw$. }

\subsection{Background and Derivation}

In the remainder of this section we will attempt to place the RTM equations
in a broader context, i.e., study how they are related to general
asymptotically flat Einstein-Maxwell fields and how from that point of view
they can be derived. We begin with the general asymptotic version of the
Einstein-Maxwell fields in a Bondi coordinate and tetrad system and see how
to specialize to the RTM equations. Later, using these results, \textit{we
will reverse the procedure and go from the RTM equations to the Bondi point
of view} in order to obtain the physical meaning of the RTM variables.

\subsubsection{Bondi coordinates and tetrad}

\textbf{\ }We start with the full set of asymptotic Maxwell and Bianchi
identities in Bondi coordinates, $(u_{B},\zeta ,\overline{\zeta })$ and
Bondi tetrad $(l,n,m,\overline{m})$\cite{Tod,Gil.Ted}$,$ (the $u_{B}$ is
already rescaled)

\underline{Maxwell Equations}

\begin{eqnarray}
&&(\phi _{0}^{0})^{{\large \cdot }}+\eth \phi _{1}^{0}-\sigma \phi
_{2}^{0}=0,  \label{phiB00cdot} \\
&&(\phi _{1}^{0})^{{\large \cdot }}+\eth \phi _{2}^{0}=0.  \label{phiB10cdot}
\end{eqnarray}

\underline{Bianchi Identities}

\begin{eqnarray}
(\psi _{0}^{0\,})^{{\large \cdot }} &=&-\eth \psi _{1}^{0}+3\sigma \psi
_{2}^{0}+3k\phi _{0}^{0}\overline{\phi }_{2}^{0},  \label{Bondi1} \\
(\psi _{1}^{0\,})^{{\large \cdot }} &=&-\eth \psi _{2}^{0}+2\sigma \psi
_{3}^{0}+2k\phi _{1}^{0}\overline{\phi }_{2}^{0},  \label{Bondi2} \\
(\psi _{2}^{0\,})^{{\large \cdot }} &=&-\eth \psi _{3}^{0}+\sigma \psi
_{4}^{0}+k\phi _{2}^{0}\overline{\phi }_{2}^{0},  \label{Bondi3} \\
\psi _{3}^{0} &=&\eth \overline{\sigma }^{{\large \cdot }},  \label{Bondi4}
\\
\psi _{4}^{0} &=&-\overline{\sigma }^{{\large \cdot \cdot }},  \label{Bondi5}
\\
\psi _{2}^{0\,}-\overline{\psi }_{2}^{0\,} &=&\overline{\eth }^{2}\sigma
-\eth ^{2}\overline{\sigma }+\sigma ^{{\large \cdot }}\overline{\sigma }-%
\overline{\sigma }^{{\large \cdot }}\sigma ,  \label{Bondi6}
\end{eqnarray}%
where
\begin{equation}
k=\frac{2G}{c^{4}}.  \label{k}
\end{equation}

From the definition of the mass aspect
\begin{equation}
\Psi =\psi _{2}^{0\,}+\eth ^{2}\overline{\sigma }+\sigma \overline{\sigma }^{%
{\large \cdot }},  \label{Psi}
\end{equation}%
we see that (\ref{Bondi6}) is equivalent to
\begin{equation}
\overline{\Psi }=\Psi .  \label{PsiReal}
\end{equation}%
and by eliminating $\psi _{2}^{0\,}$ in terms of $\Psi ,$ Eq.(\ref{Bondi3})
becomes the \textit{very important Bondi mass loss equation}
\begin{equation}
\Psi ^{{\large \cdot }}=\sigma ^{{\large \cdot }}(\overline{\sigma })^{%
{\large \cdot }}+k\phi _{2}^{0}\overline{\phi }_{2}^{0}.  \label{evolution}
\end{equation}

\subsubsection{Asymptotically shear-free tetrad}

These relations take a new form after using the tetrad transformation, (the
asymptotic null rotation around $n),$
\begin{eqnarray}
l^{*a} &=&l^{a}+b\overline{m}^{a}+\overline{b}m^{a}+b\overline{b}n^{a},
\label{NullRot} \\
m^{*a} &=&m^{a}+bn^{a},  \notag \\
n^{*a} &=&n^{a},  \notag \\
b &=&-L/r+O(r^{-2}),  \notag \\
L &=&L(u_{B},\zeta ,\overline{\zeta }),  \notag
\end{eqnarray}
with
\begin{eqnarray}
\phi _{0}^{0} &=&\phi _{0}^{*0}+2L\phi _{1}^{*0}+L^{2}\phi _{2}^{*0},
\label{BtoTphi00} \\
\phi _{1}^{0} &=&\phi _{1}^{*0}+L\phi _{2}^{*0},  \label{BtoTphi01} \\
\phi _{2}^{0} &=&\phi _{2}^{*0},  \label{BtoTphi02}
\end{eqnarray}
\begin{eqnarray}
\psi _{0}^{0} &=&\psi _{0}^{*0}+4L\psi _{1}^{*0}+6L^{2}\psi
_{2}^{*0}+4L^{3}\psi _{3}^{*0}+L^{4}\psi _{4}^{*0},  \label{BtoTpsi00} \\
\psi _{1}^{0} &=&\psi _{1}^{*0}+3L\psi _{2}^{*0}+3L^{2}\psi
_{3}^{*0}+L^{3}\psi _{4}^{*0},  \label{BtoTpsi01} \\
\psi _{2}^{0} &=&\psi _{2}^{*0}+2L\psi _{3}^{*0}+L^{2}\psi _{4}^{*0},
\label{BtoTpsi02} \\
\psi _{3}^{0} &=&\psi _{3}^{*0}+L\psi _{4}^{*0},  \label{BtoTpsi03} \\
\psi _{4}^{0} &=&\psi _{4}^{*0}.  \label{BtoTpsi04}
\end{eqnarray}

{\bf Remark 7.} {\it Note that the ( $^{*}$) used here has nothing
to do with the ( $^{*}$) from the rescaling of $V$ and $\tau $ in
section 3. }

First, we obtain a new asymptotic shear\cite{Aronson,Ftprints} $\sigma
^{\ast }$ for the null vector $l^{\ast \,a},$ that is given by
\begin{equation*}
\sigma ^{\ast }=\sigma -(\eth L+LL^{{\large \cdot }})\text{.}
\end{equation*}%
The requirement of a vanishing $\sigma ^{\ast }$ leads to the differential
equation for the determination of $L,$%
\begin{equation}
\eth L+LL^{{\large \cdot }}=\sigma .  \label{shearfree}
\end{equation}

Using such an $L$, with Eq.(\ref{NullRot}), the asymptotic Maxwell and
Bianchi Identities are changed to equivalent equations (just with a \textit{%
\ \ changed tetrad)} but with the \textit{same Bondi coordinates. } Now
however $l^{*a}$ defines an asymptotically shear-free and, in general
twisting, null vector. Shortly we consider a special case where the twist
vanishes.

\qquad \underline{Maxwell Equations}
\begin{eqnarray}
(\phi _{0}^{\ast 0})^{{\large \cdot }}+\eth \phi _{1}^{\ast 0}+2L^{{\large %
\cdot }}\phi _{1}^{\ast 0}+L(\phi _{1}^{\ast 0})^{{\large \cdot }} &=&0,
\label{ME1T} \\
(\phi _{1}^{\ast 0})^{{\large \cdot }}+\eth \phi _{2}^{\ast 0}+(L\phi
_{2}^{\ast 0})^{{\large \cdot }} &=&0.  \label{ME2T}
\end{eqnarray}

\qquad \underline{Bianchi Identities}

\begin{eqnarray}
(\psi _{0}^{\ast 0\,})^{{\large \cdot }} &=&-\eth \psi _{1}^{\ast 0}-L(\psi
_{1}^{\ast 0\,})^{{\large \cdot }}-4L^{{\large \cdot }}\psi _{1}^{\ast
0}+3k\phi _{0}^{\ast 0}\overline{\phi }_{2}^{\ast 0},  \label{twisting1} \\
(\psi _{1}^{\ast 0\,})^{{\large \cdot }} &=&-\eth \psi _{2}^{\ast 0}-L(\psi
_{2}^{\ast 0\,})^{{\large \cdot }}-3L^{{\large \cdot }}\psi _{2}^{\ast
0}+2k\phi _{1}^{\ast 0}\overline{\phi }_{2}^{\ast 0},  \label{twisting2} \\
(\psi _{2}^{\ast 0\,})^{{\large \cdot }} &=&-\eth \psi _{3}^{\ast 0}-L(\psi
_{3}^{\ast 0\,})^{{\large \cdot }}\,-2L^{{\large \cdot }}\psi _{3}^{\ast
0}+k\phi _{2}^{\ast 0}\overline{\phi }_{2}^{\ast 0},  \label{twisting3} \\
\psi _{3}^{\ast 0} &=&\eth (\overline{\sigma })^{{\large \cdot }}+L(%
\overline{\sigma })^{{\large \cdot \cdot }},  \label{twisting4} \\
\psi _{4}^{\ast 0} &=&-\overline{\sigma }^{{\large \cdot \cdot }},
\label{twisting5*} \\
\psi _{2}^{\ast 0}-\overline{\psi }_{2}^{\ast 0} &=&\overline{\eth }%
^{2}\sigma -\eth ^{2}\overline{\sigma }+\overline{\sigma }\sigma ^{{\large %
\cdot }}-\sigma \overline{\sigma }^{{\large \cdot }}  \label{Bondi6**} \\
&&+2\overline{L}\,\overline{\eth }(\sigma )^{{\large \cdot }}-2L\text{ }\eth
(\overline{\sigma })^{{\large \cdot }}+\overline{L}^{2}\sigma ^{{\large %
\cdot \cdot }}-L^{2}\overline{\sigma }^{{\large \cdot \cdot }},  \notag
\end{eqnarray}%
where now we treat $L(u_{B},\zeta ,\overline{\zeta })$ as the basic
asymptotic data and $\sigma $ is \textit{defined} by
\begin{equation}
\sigma \equiv \eth L+LL^{{\large \cdot }}.  \label{twisting6}
\end{equation}%
By using (\ref{BtoTpsi02}) in (\ref{Psi}) we find that
\begin{equation}
\Psi =\psi _{2}^{\ast 0}+2L\eth (\overline{\sigma })^{{\large \cdot }}+L^{2}%
\overline{\sigma }^{{\large \cdot \cdot }}+\eth ^{2}\overline{\sigma }%
+\sigma \overline{\sigma }^{{\large \cdot }},  \label{Psi*}
\end{equation}%
so that (\ref{Bondi6**}) is again equivalent to
\begin{equation}
\Psi -\overline{\Psi }=0.  \label{twisting8}
\end{equation}%
Note also that only the first three of Eqs.(\ref{twisting1})-(\ref%
{twisting5*}) have a dynamic content, Eqs.(\ref{twisting4}) and (\ref%
{twisting5*}) are just definitions that can be inserted into Eq.(\ref%
{twisting3}). This fact will be used later.

\subsubsection{Type II metrics in Bondi coordinates}

Up to now the two sets of equations, with the Bondi tetrad and the
shear-free tetrad, are completely equivalent. We now make our major
condition that is a severe restriction on the solutions. We assume that we
are dealing with\textit{\ algebraically special type\ II Einstein-Maxwell
metrics }with a non-twisting pnv\textbf{, }$l^{*a}.$ This translates $%
immediately$ into the statement that $\psi _{0}^{*0\,}=\psi _{1}^{*0\,}=0$
and that automatically, from Eq.(\ref{twisting1}), we have that $\phi
_{0}^{*0}=0.$ The non-twisting assumption has not yet been used.

Using this the Maxwell/Bianchi Identities become

\begin{eqnarray}
\eth \phi _{1}^{\ast 0}+2L^{{\large \cdot }}\phi _{1}^{\ast 0}+L(\phi
_{1}^{\ast 0})^{{\large \cdot }} &=&0,  \label{m1} \\
(\phi _{1}^{\ast 0})^{{\large \cdot }}+\eth \phi _{2}^{\ast 0}+(L\phi
_{2}^{\ast 0})^{{\large \cdot }} &=&0,  \label{m2}
\end{eqnarray}

\begin{eqnarray}
0 &=&-\eth \psi _{2}^{\ast 0}-L(\psi _{2}^{\ast 0\,})^{{\large \cdot }}-3L^{%
{\large \cdot }}\psi _{2}^{\ast 0}+2k\phi _{1}^{\ast 0}\overline{\phi }%
_{2}^{\ast 0},  \label{gr1} \\
(\psi _{2}^{\ast 0\,})^{{\large \cdot }} &=&-\eth \psi _{3}^{\ast 0}-L(\psi
_{3}^{\ast 0\,})^{{\large \cdot }}\,-2L^{{\large \cdot }}\psi _{3}^{\ast
0}+k\phi _{2}^{\ast 0}\overline{\phi }_{2}^{\ast 0},  \label{gr2} \\
\Psi &=&\psi _{2}^{\ast 0}+2L\eth(\overline{\sigma })^{{\large \cdot }%
}+L^{2}(\overline{\sigma })^{{\large \cdot \cdot }}+\eth ^{2}\overline{%
\sigma }+\sigma (\overline{\sigma })^{{\large \cdot }}=\overline{\Psi },
\label{gr3}
\end{eqnarray}%
with, as mentioned above, the defining relations

\begin{eqnarray}
\psi _{3}^{\ast 0} &\equiv &\eth (\overline{\sigma })^{{\large \cdot }}+L(%
\overline{\sigma })^{{\large \cdot \cdot }}  \label{psi3} \\
\sigma  &\equiv &\eth L+LL^{{\large \cdot }}.  \label{SIGMA}
\end{eqnarray}

They are a set of coupled equations for the variables $(L,\phi
_{1}^{*0},\phi _{2}^{*0},\psi _{2}^{*0}).$

The shear is no longer free data - it follows from knowledge of $L,$ via Eq.(%
\ref{SIGMA}).

\subsubsection{Transformation to NU/RT coordinates}

By transforming to a new set of coordinates at $\mathfrak{I}^{+}$, namely
NU/RT coordinates, [with the same tetrad], we obtain a great simplification
in these equations.

The transformation is to be given by\cite{Ftprints,Gil.Ted}
\begin{equation}
u_{B}=X(\tau ,\zeta ,\overline{\zeta }),  \label{X(tau)}
\end{equation}%
($\zeta $ unchanged) where $X(\tau ,\zeta ,\overline{\zeta })$ is obtained
from the following construction, based on the (assumed) known $L(u_{B},\zeta
,\overline{\zeta }):$ First the (CR\cite{CR}) function,
\begin{subequations}
\begin{equation}
\tau =T(u_{B},\zeta ,\overline{\zeta }),  \label{T}
\end{equation}%
is found by solving the differential equation
\end{subequations}
\begin{equation}
\eth T+LT^{\,{\large \cdot }}=0,  \label{CR}
\end{equation}%
and then by inverting it, i.e., by algebraically solving for $u_{B},$ the $%
X(\tau ,\zeta ,\overline{\zeta })$ is constructed.

{\bf Remark 8.} {\it
In the more general situation of twisting type II congruences, the $T$ and $%
X $ are in general complex-valued analytic functions, however for
the (non-twisting) Robinson-Trautman case they are real. }

Now applying this transformation, $u_{B}\Rightarrow \tau ,$ to Eqs.(\ref{m1}%
)-(\ref{gr3}), (a lengthy process), we obtain our final version of the
Maxwell/Bianchi Identities in the new tetrad basis with the NU/RT
coordinates, $(\tau ,\zeta ,\overline{\zeta }).$ The variable $L(u_{B},\zeta
,\overline{\zeta })$ is replaced by the new basic variable
\begin{equation}
V(\tau ,\zeta ,\overline{\zeta })\equiv X^{\prime }(\tau ,\zeta ,\overline{
\zeta }).
\end{equation}

{\bf Remark 9.} {\it Frequently, in the application of this
transformation, we have used the easily derived relations
\begin{eqnarray}
L &=&-\frac{\eth T}{T^{\,\,{\large \cdot }}}=\eth _{(\tau )}X,  \label{L} \\
L^{{\large \cdot }} &=&V^{-1}\text{ \dh }_{(\tau )}V,  \label{Ldot} \\
\sigma  &=&\eth L+LL^{{\large \cdot }}=\eth _{(\tau )}^{2}X,  \label{sigma2}
\end{eqnarray}%
where \dh $_{(\tau )}$ means the \dh -operator but holding $\tau $
constant as opposed to holding $u_{B}$ constant. }

\subsubsection{\textbf{Maxwell/RT in NU/RT coordinates}:}

\begin{eqnarray}
\eth_{(\tau )}[V^{2}\phi _{1}^{*0}] &=&0\Rightarrow \phi _{1}^{*0}\equiv
\frac{q}{2}V^{-2},  \label{m1*} \\
\phi _{1}^{*0\prime }+\eth_{(\tau )}[V\phi _{2}^{*0}] &=&0,  \label{m2*}
\end{eqnarray}

\begin{eqnarray}
-\eth_{(\tau )}\chi &=&2kV^{3}\phi _{1}^{*0}\overline{\phi }_{2}^{*0},
\label{gr1*} \\
\chi &=&-V^{3}\psi _{2}^{*0},  \label{gr2*} \\
3\chi \frac{V^{\prime }}{V}-\chi ^{\prime }+V^{3}\{\eth_{(\tau )}^{2}%
\overline{\eth}_{(\tau )}^{2}V-V^{-1}\overline{\eth}_{(\tau )}^{2}V\cdot
\eth_{(\tau )}^{2}V\} &=&kV^{4}\phi _{2}^{*0}\overline{\phi }_{2}^{*0},
\label{gr3*}
\end{eqnarray}
with ($\chi ,V$, $\psi _{2}^{*0})$ all real. The mass aspect simplifies,
after a lengthy calculation, to

\begin{equation}
\Psi =\psi _{2}^{*0}+\eth_{(\tau )}^{2}\overline{\eth}_{(\tau )}^{2}X=%
\overline{\Psi }.  \label{PSI**}
\end{equation}

These equations are precisely the same as those given earlier, Eqs.(\ref{max}
)-(\ref{RT2}). They had been obtained\cite{Tod}, with some notation changes,
by direct integration of the NP equations assuming that $\psi
_{0}^{*0\,}=\psi _{1}^{*0\,}=\phi _{0}^{*0}=0$ with a real `divergence' $%
\rho .$

\subsection{Approximate solutions}

As we mentioned (and is easily seen) the RTM equations are quite non-linear
and difficult to solve. We have resorted to an approximation scheme based on
the idea that the dominant term in gravitational radiation is from the
quadrupole. We thus assume that all quantities are to be expanded in
spherical harmonics including terms from $l=0,1,2$ with nothing higher.
Essentially we are approximating off the Reissner-Nordstom metric which we
take as zeroth order. We consider terms up to quadratic in the deviation
from the zeroth order. Due to the non-linearities there are frequent
products of the $l=0,1,2$ harmonics which are then decomposed via the
Clebsch-Gordon expansions (Sec. II). When these expansions involve harmonics
with $l>2$ these higher $l$ terms will be omitted. In a few instances where
the $l=2$ harmonics or the quadratic terms do not appear to be important, we
will omit them. This occurs for the Maxwell fields, which appear in the
gravitational equations only as quadratic terms. These $l=2$ could easily be
restored but to keep the equations tractable their omission seems to be
reasonable. Where this occurs, we will point it out.

We use the following notation (remembering the rescaling of $\tau $) for the
harmonic expansions of the relevant variables:
\begin{eqnarray}
V &=&\frac{v_{0}}{\sqrt{2}}-\frac{1}{2}v^{i}Y_{1i}^{0}+v^{ij\,}Y_{2ij}^{0}=%
\frac{\sqrt{2}+\delta v_{0}}{\sqrt{2}}-\frac{1}{2}v^{i}Y_{1i}^{0}+v^{ij%
\,}Y_{2ij}^{0},  \label{Vh} \\
V^{\prime } &=&\frac{\delta v_{0}^{\prime }}{\sqrt{2}}-\frac{1}{2}
v^{i\,\prime }Y_{1i}^{0}+v^{ij\,\prime }Y_{2ij}^{0},  \label{V'h} \\
X &=&\int^{\tau }Vd\tau =\frac{\sqrt{2}\tau +\int \delta v_{0}d\tau }{\sqrt{2%
}}-\frac{1}{2}\xi ^{i}Y_{1i}^{0}+\xi ^{ij}Y_{2ij}^{0},  \label{Xh} \\
\xi ^{i\prime } &=&v^{i},\qquad \xi ^{ij\prime }=v^{ij},  \label{v^ih} \\
V\phi _{2}^{*0} &=&\Phi ^{i}Y_{1i}^{-1}+\Phi ^{ij}Y_{2ij}^{-1},  \label{phih}
\\
\chi &=&\chi _{0}+\delta \chi _{0}+\chi ^{i}Y_{1i}^{0}+\chi ^{ij}Y_{2ij}^{0}.
\label{xi}
\end{eqnarray}

Our first differential equation, Eq.(\ref{max}),
\begin{equation}
\phi _{1}^{*0\prime }+\eth_{(\tau )}[V\phi _{2}^{*0}]=0,  \label{max*}
\end{equation}
becomes, using the properties of edth,

\begin{eqnarray}
-(\frac{q}{2}V^{-2})^{\prime } &=&qV^{-3}V^{\prime }=\eth_{(\tau )}[V\phi
_{2}^{*0}]=\eth_{(\tau )}[\Phi ^{i}Y_{1i}^{-1}+\Phi ^{ij}Y_{2ij}^{-1}],
\label{max2} \\
qV^{-3}V^{\prime } &=&\Phi ^{i}Y_{1i}^{0}+\Phi ^{ij}Y_{2ij}^{0}.  \notag
\end{eqnarray}

From the Taylor expansion of $V^{-3},$ with (\ref{V'h}), we have the
solutions,
\begin{eqnarray}
V\phi _{2}^{*0} &=&-q[\sqrt{2}v^{i\,\prime }-\frac{18\cdot 4}{5}
(v^{j}v^{ij})^{\prime }]Y_{1i}^{-1}  \label{sol.phi} \\
&&+q\left\{ 2\sqrt{2}v^{ji\prime }-Sym.T.F.[v^{j}v^{i\prime }+\frac{(24)(12)%
}{7}v^{jl}v^{il\prime }]\right\} Y_{2ij}^{-1}.  \notag
\end{eqnarray}
$Sym.T.F.$ means symmetrized and trace free part of the expression and $\chi
_{0}$ is constant.

The linearized Maxwell field is

\begin{eqnarray}
\phi _{0}^{*0} &=&0,  \label{dipolefield} \\
\phi _{1}^{*0} &=&q[1+\sqrt{2}v^{i}Y_{1i}^{0}-2\sqrt{2}v^{ki}Y_{2ki}^{0}],
\label{dipole2} \\
\phi _{2}^{*0} &=&-2q[v^{i\prime }Y_{1i}^{-1}-2v^{ij\prime }Y_{2ij}^{-1}].
\label{dipole3}
\end{eqnarray}

(See appendix A for the solution up to second order with the $l=2$
harmonics.)

Using Eqs.(\ref{coulomb}), (\ref{xi}) and (\ref{sol.phi}) in Eq.(\ref{RT}),
i.e.,
\begin{equation}
\eth_{(\tau )}\chi +2kV^{3}\phi _{1}^{*0}\overline{\phi }_{2}^{*0}=0,
\label{RT**}
\end{equation}
we find that
\begin{eqnarray}
\text{ }\chi ^{i} &=&-\frac{kq^{2}}{2}(\sqrt{2}v^{i\prime }-\frac{72}{5}
\{v^{j}v^{ij}\}^{\prime }),  \notag \\
\chi ^{ij} &=&\frac{kq^{2}\sqrt{2}}{3}v^{ji\prime }-\frac{kq^{2}}{6}Sym.T.F.%
{\large (}v^{j}v^{i\prime }+\frac{288}{7}v^{jl}v^{il\prime }{\large )}.
\notag
\end{eqnarray}

Finally the last and most complicated equation, (\ref{RT2}), is studied
after inserting $\phi _{2}^{*0}$ and $\chi $ from Eqs.(\ref{sol.phi}) and (%
\ref{xi}). We find, after considerable effort, that the evolution is given by

\begin{eqnarray}
(\delta \chi _{0})^{\prime } &=&3\chi _{0}v^{i}v^{i\prime }+\frac{48(18)}{5}
v^{ij}v^{ij}  \label{evol1} \\
&&+kq^{2}\{v^{i\prime }v^{i\prime }-\frac{1}{3}[v^{i}v^{i\prime }]^{\prime }-%
\frac{16}{5}[v^{ij\prime }v^{ij}]^{\prime }+\frac{48}{5}v^{ji\prime
}v^{ij\prime }\},  \notag \\
\chi _{0}v^{i\prime } &=&-\frac{96\sqrt{2}}{5}v^{j}v^{ij}+q^{2}k\{\frac{1}{3}
[v^{i\prime \prime }-\frac{36\sqrt{2}}{5}(v^{j}v^{ij})^{\prime \prime }
\label{evol2*} \\
&&+\frac{12\sqrt{2}}{5}(v^{ij}v^{j\prime })^{\prime }]+\frac{4\sqrt{2}}{15}[
v^{ij\prime \prime }v^{j}-9v^{ij\prime }v^{j\prime }-3v^{ji\prime
}v^{j\prime }]\},  \notag \\
3\chi _{0}v^{ij\prime } &=&-6v^{ij}-Sym.T.F.[\frac{48\cdot 30}{7\sqrt{2}}
v^{il}v^{jl}]+\frac{kq^{2}}{3}v^{ij\prime \prime }  \label{evol3} \\
&&+Sym.T.F.[\frac{16kq^{2}}{7\sqrt{2}}(v^{jl\prime
}v^{il}-3v^{jl}v^{il\prime \prime }-\frac{21}{64}v^{i\prime }v^{j\prime }-%
\frac{11}{2}v^{jl\prime }v^{il\prime })].  \notag
\end{eqnarray}

Equations (\ref{evol1}), (\ref{evol2*}) and (\ref{evol3}) constitute our
final set of (ordinary) differential equations within our approximation
scheme for the RTM equations, while Eqs.(\ref{xi}) with (\ref{dipolefield})
and (\ref{coulomb}) yield the asymptotic physical Maxwell and Weyl fields.
Even with our `heavy' approximations it is not easy to understand them.
However, in the next section, where we transform the RTM equations to Bondi
coordinates, we can see meaning to these equations from physical insight.
This physical insight might even help in understanding the mathematical
structures involved.

\section{The Bondi Energy-Momentum Four-Vector}

In order to construct the Bondi energy-momentum four-vector and its
evolution we reverse the coordinate transformation that took us from Bondi
coordinates to the NU/RT coordinates that was given earlier in Eqs.(\ref%
{X(tau)}) and (\ref{T}). We then express the mass aspect in terms of the RTM
variables.

\subsection{Transformation to Bondi coordinates}

In order to go from the present NU/RT coordinates, ($\tau ,\zeta ,\overline{%
\zeta }$) on null infinity, ($\mathfrak{I}^{+}$), to the Bondi coordinates, (%
$u_{B},\zeta ,\overline{\zeta }$), we simply use Eq.(\ref{X(tau)}) with its
approximation

\begin{eqnarray}
u_{B} &=&X(\tau ,\zeta ,\overline{\zeta })  \label{NU>Bondi} \\
&=&\int^{\tau }V(\tau ^{\prime },\zeta ,\overline{\zeta })d\tau ^{\prime }=%
\frac{\xi ^{0}(\tau )}{\sqrt{2}}-\frac{1}{2}\xi ^{i}(\tau )Y_{1i}^{0}+\xi
^{ij}(\tau )Y_{2ij}^{0},  \label{NU>Bondi2}
\end{eqnarray}
with ($\zeta ,\overline{\zeta })$ unchanged. We need to find the inversion
of (\ref{NU>Bondi2}), i.e., we must find the approximate $\tau
=T(u_{B},\zeta ,\overline{\zeta })$.

{\bf Remark 10.} {\it If \textit{exact solutions}, $V(\tau ,\zeta
,\overline{\zeta }),$ of the RTM equations were known, then the
\textit{exact transformation} to Bondi coordinates at
$\mathfrak{I}^{+}$ would be given by
\begin{equation*}
u_{B}=X(\tau ,\zeta ,\overline{\zeta })=\int^{\tau }V(\tau ^{\prime },\zeta ,%
\overline{\zeta })d\tau ^{\prime }\text{,}
\end{equation*}
with an additive arbitrary supertranslation. By ignoring it, we
implicitly have chosen a specific Bondi frame. }

{\bf Remark 11.} {\it
In this section we will be returning to the unrescaled $\tau ,$ so that $%
\tau \Rightarrow \tau /\sqrt{2},$ (see Eq.(\ref{rescale1}), (\ref{Vh}) and (%
\ref{Xh})). }

Before finding $T(u_{B},\zeta ,\overline{\zeta }),$ we point out that $%
\partial _{u_{B}}T\equiv T^{\cdot }=V^{-1}$ since by differentiating (\ref%
{NU>Bondi})
\begin{equation}
1=V\frac{d\tau }{du_{B}}=VT^{\,\,{\large \cdot }}.  \label{V-1}
\end{equation}

The relation $\tau =T(u_{B},\zeta ,\overline{\zeta })$ can be approximated
in the following manner: from Eq.(\ref{Vh}), we have that
\begin{equation}
\xi ^{0}(\tau )=\tau +\delta \xi ^{0}(\tau ),  \label{x0.2}
\end{equation}
so Eq.(\ref{NU>Bondi}) can be rewritten, with
\begin{equation}
w=\sqrt{2}u_{B},  \label{w}
\end{equation}
as
\begin{eqnarray}
\tau &=&w+F(\tau ,\zeta ,\overline{\zeta }),  \label{F} \\
F(\tau ,\zeta ,\overline{\zeta }) &\equiv &-\delta \xi ^{0}(\tau )+\frac{%
\sqrt{2}}{2}\xi ^{i}(\tau )Y_{1i}^{0}-\sqrt{2}\xi ^{ij}(\tau )Y_{2ij}^{0}.
\notag
\end{eqnarray}

This can be iterated with
\begin{equation}
\tau =w  \label{zeroth}
\end{equation}
as the zeroth order iterate. The first iterate of Eq.(\ref{F}) is then
\begin{eqnarray}
\tau &=&w+F(w,\zeta ,\overline{\zeta }),  \label{1stIterate} \\
\tau &=&w-\delta \xi ^{0}(w)+\frac{\sqrt{2}}{2}\xi ^{i}(w)Y_{1i}^{0}-\sqrt{2}
\xi ^{ij}(w)Y_{2ij}^{0},  \label{1stIterate*}
\end{eqnarray}
with the second iterate
\begin{equation}
\tau =w+F{\large (}w+F(w,\zeta ,\overline{\zeta }),\zeta ,\overline{\zeta }%
{\large )}\text{ }\approx \text{ }w+F(w,\zeta ,\overline{\zeta })+F\partial
_{w}F,  \label{2ndIt}
\end{equation}
with higher iterates (though unpleasant and not needed) easily found. We
then have that
\begin{equation*}
T^{\,\,{\large \cdot }}=\sqrt{2}\partial _{w}T=\sqrt{2}\mathbf{(}1+\partial
_{w}F(w)+\partial _{w}^{2}F(w)\cdot F(w)+\partial _{w}F(w)\cdot \partial
_{w}F(w)),
\end{equation*}
or, using the Clebsch-Gordon expansions and omitting the argument $w$,

\begin{eqnarray}
T^{\,\,{\large \cdot }} &=&\sqrt{2}{\large (}1+\frac{1}{3}\{v^{i}\xi
^{i}\}^{\prime }-\delta v_{0}+\frac{48}{5}[\xi ^{ij}v^{ij\prime
}+v^{ij}v^{ij}]  \label{Tdot} \\
&&+[\frac{\sqrt{2}}{2}v^{i}-\frac{12}{5}\{v^{ij\prime }\xi ^{j}+v^{j\prime
}\xi ^{ij}+2v^{ij}v^{j}\}]Y_{1i}^{0}  \notag \\
&&+\frac{1}{6}\{v^{i\prime }\xi ^{k}+v^{i}v^{k}-6\sqrt{2}v^{ki}+\frac{288}{7}%
[\xi ^{kl}v^{il\prime }+v^{kl}v^{il}]\}Y_{2ki}^{0}{\large ).}  \notag
\end{eqnarray}

This result plays an important role in determining the mass aspect and the
Bondi four-vector.

{\bf Remark 12.} {\it
We have been using the notation $(^{\prime })\equiv \partial _{\tau }$ and $%
(^{\cdot })\equiv \partial _{u_{B}}.$ In the following where $w,$ from Eq.(%
\ref{w}), is used in the same functional expressions that had used
$\tau ,$ we will \textit{continue to use the notation } }
\begin{equation*}
(^{\prime })\equiv \partial _{w}=\frac{1}{\sqrt{2}}(^{{\large \cdot }})\text{
.}
\end{equation*}

\subsection{The Mass Aspect}

Our task is to take the mass aspect from Eq.(\ref{PSI**})
\begin{eqnarray}
\Psi &=&\overline{\Psi }=\psi _{2}^{*0}+\eth_{(\tau )}^{2}\overline{\eth}%
_{(\tau )}^{2}X  \label{PSItau} \\
&=&-V^{-3}\chi +\eth_{(\tau )}^{2}\overline{\eth}_{(\tau )}^{2}X,  \notag
\end{eqnarray}
which is a function of ($\tau ,\zeta ,\overline{\zeta }$) and reexpress it,
via $\tau =T((u_{B},\zeta ,\overline{\zeta }),$ in Bondi coordinates.

All the terms on the right side of (\ref{PSItau}) are known functions of ($%
\chi _{0},\delta v^{0},v^{i},v^{ij},\delta \xi ^{0},$ $\xi ^{i},\xi ^{ij}$),
which in turn are functions of $(\tau ,\zeta ,\overline{\zeta }).$ These
known functions are substituted into Eq.(\ref{PSItau}) and then $\tau $ is
replaced by the first iterate, Eq.(\ref{1stIterate*}). \{The first iterate
is sufficient since all the relevant quantities are already 1$^{st}$ order.\}

After another very lengthy calculation, using the third power of $%
T^{\,\,\cdot },$ i.e., Eq.(\ref{Tdot}), and frequent use of Clebsch-Gordon
expansions, we finally find (including only the zeroth and first harmonic
terms, the only ones needed for the Bondi four-vector)
\begin{eqnarray}
\Psi (w,\zeta ,\overline{\zeta }) &=&\Psi ^{(0)}+\Psi ^{(1i)}Y_{1i}^{0},
\label{MassAspect4} \\
\Psi ^{(0)} &=&-2\sqrt{2}{\large \{}\chi _{0}[1+\frac{1}{2}
v^{i}v^{i}+v^{i\prime }\xi ^{i}+\frac{144}{5}(\xi ^{ij}v^{ij\prime
}+2v^{ij}v^{ij})]+\frac{288}{5}\xi ^{ij}v^{ij}  \notag \\
&&+\delta \chi _{0}-q^{2}k[\frac{1}{3}v^{i\prime \prime }\xi ^{i}+\frac{16}{5%
}\xi ^{ij}v^{ij\prime \prime }+\frac{48}{5}v^{ij}v^{ij\prime }+v^{i\prime
}v^{i}]{\large \},}  \notag \\
\Psi ^{(1i)} &=&-6{\large \{}\chi _{0}[v^{i}-\frac{12\sqrt{2}}{5}
(v^{ij\prime }\xi ^{j}+v^{j\prime }\xi ^{ij}+4v^{ij}v^{j})]-\frac{48}{5\sqrt{%
2}}v^{ij}\xi ^{j}  \notag \\
&&+\frac{\sqrt{2}q^{2}k}{3}[-\frac{1}{\sqrt{2}}v^{i\prime }+\frac{48}{5}
(v^{j}v^{ij})^{\prime }+\frac{12}{5}\xi ^{ij}v^{j\prime \prime }+\frac{4}{5}
v^{ij\prime \prime }\xi ^{j}+\frac{24}{5}v^{ij}v^{j\prime }]{\large \}.}
\notag
\end{eqnarray}

{\bf Remark 13.} {\it As already pointed out the already defined
$w=$ $\sqrt{2}u_{B}$ is really the unrescaled time coordinate,
Eq.(\ref{Bondi}), that was originally used. In what follows we
replace $w$ by $cw,$ i.e., }

\begin{eqnarray*}
w &\Rightarrow &cw, \\
(^{\prime }) &\Rightarrow &c^{-1}(^{\prime })\text{.}
\end{eqnarray*}

Since the relationship\cite{Gil.Ted,Carlos.Ted} between the mass aspect and
the Bondi energy-momentum four-vector, ($Mc,P^{i})$ is given (with $%
M=M_{0}+\delta M_{0}$ and $M_{0}=const.$) by
\begin{equation}
\Psi (w,\zeta ,\overline{\zeta })=-\frac{2\sqrt{2}G}{c^{2}}M-\frac{6G}{c^{3}}
P^{i}Y_{1i}^{0}+...,  \label{PSI.M.P}
\end{equation}
then by comparison with Eq.(\ref{MassAspect4}), we can make the physical
identifications: with

\begin{equation*}
\chi _{0}=\frac{G}{c^{2}}M_{0}
\end{equation*}
we have that
\begin{eqnarray}
\text{ }M &=&M_{0}[1+\frac{v^{i2}}{2c^{2}}+\frac{v^{i\prime }\xi ^{i}}{c^{2}}
+\frac{144}{5c^{2}}(\xi ^{ij}v^{ij\prime }+2v^{ij}v^{ij})]+\frac{288c}{5G}
\xi ^{ij}v^{ij}  \label{M*} \\
&&-\frac{2q^{2}}{c^{5}}[v^{i\prime }v^{i}+\frac{1}{3}v^{i\prime \prime }\xi
^{i}+\frac{16}{5}\xi ^{ij}v^{ij\prime \prime }+\frac{48}{5}v^{ij}v^{ij\prime
}],  \notag \\
P^{i} &=&M_{0}v^{i}-\frac{2q^{2}}{3c^{3}}v^{i\prime }-\frac{12\sqrt{2}}{5c}
M_{0}(v^{ij\prime }\xi ^{j}+v^{j\prime }\xi ^{ij}+4v^{ij}v^{j})-\frac{24%
\sqrt{2}c^{2}}{5G}v^{ij}\xi ^{j}  \label{P^i} \\
&&+\frac{8\sqrt{2}q^{2}}{15c^{4}}[12(v^{j}v^{ij})^{\prime }+3\xi
^{ij}v^{j\prime \prime }+v^{ij\prime \prime }\xi ^{j}+6v^{ij}v^{j\prime }].
\notag
\end{eqnarray}

\subsection{The Bondi Energy-Momentum 4-Vector}

The Bondi energy-momentum four-vector,
\begin{equation}
P^{a}=(Mc,P^{i}),  \label{P^a}
\end{equation}
which has been extracted from the asymptotic Weyl tensor has been known for $%
almost$ 50 years. It transforms as a Lorentzian four-vector under the
Lorentz subgroup of the BMS group. It satisfies the beautiful Bondi
energy-momentum loss theorem. But, to our knowledge, it never before had a
kinematic relationship with anything that resembled a velocity.

We see that the variable $v^{i}$ that was introduced formally, in Sec. IV,
as the $l=1$ harmonic component in $V(\tau ,\zeta ,\overline{\zeta })$ now
plays the kinematic role of a velocity. We see that the mass has the kinetic
energy term
\begin{equation}
M_{0}c^{2}(1+\frac{1}{2}\frac{v^{2}}{c^{2}})\approx \frac{M_{0}c^{2}}{\sqrt{%
1-\frac{v^{2}}{c^{2}}}},  \label{KE}
\end{equation}
while the momentum, at lowest order, is just
\begin{equation}
P^{i}=M_{0}v^{i}.  \label{Mv}
\end{equation}

Though most of the remaining terms in both $M$ and $P^{i}$ come from the $%
l=2 $ (quadrupole) terms and remain to be understood, the several terms
involving the $l=1$ quantities (both linear and quadratic) are interesting.
In $M$ there is the anomalous term
\begin{equation*}
M_{0}v^{i\prime }\,\xi ^{i},
\end{equation*}
which can be interpreted as Force $x$ displacement, the work done on the
total system moving it to the `position' $\xi .$ The two quadratic terms
appear to be new, a contribution to the mass from an accelerated massive
charge. This appears to be a genuine prediction from this work, though the
effect is probably too small ever to be measured.

The second term in $P^{i}$, i.e.,
\begin{equation}
-\frac{2}{3c^{3}}q^{2}v^{i\,\prime },  \label{RadReact}
\end{equation}
is quite well known. In the equations of motion, obtained below, from the
energy-momentum loss theorem, the derivative of this term is precisely the
(electromagnetic) radiation reaction force with the correct numerical
factors.

The question is in what `space' is this motion taking place? We saw in Eq.(%
\ref{NU>Bondi2}), i.e.,

\begin{eqnarray}
u_{B} &=&X(\tau ,\zeta ,\overline{\zeta })=  \label{NU>Bondi5} \\
&=&\frac{\xi ^{0}(\tau )}{\sqrt{2}}-\frac{1}{2}\xi ^{i}(\tau )Y_{1i}^{0}+\xi
^{ij}(\tau )Y_{2ij}^{0}+...,  \notag
\end{eqnarray}
that we omitted an arbitrary additive (integration) term, coming from the
four-parameter Poincar\'{e} subgroup of the BMS group, of the form

\begin{equation}
\xi _{0}^{a}l_{a}=\frac{\xi _{0}^{0}}{\sqrt{2}}-\frac{1}{2}\xi
_{0}^{i}Y_{1i}^{0},  \label{translation}
\end{equation}
where $\xi _{0}^{a}$ is an arbitrary (constant) translation. We can thus
interpret the $\xi ^{a}(\tau )$ as motion in this four dimensional space and
\textit{define it as the center of mass motion} of the total system with $%
v^{a}(\tau )=\xi ^{a\prime }(\tau )$ as the center of mass velocity.

{\bf Remark 14.} {\it Though this issue will not be pursued here,
there is an alternate space where the motion or world-line can be
considered as taking place. Associated with each RTMaxwell solution
there is defined a complex four-dimensional manifold referred to as
H-space\cite{Ftprints,H-space1,H-space2}. Our world-line can be
interpreted as a curve in H-space. }

\subsection{Energy-Momentum Loss Theorem}

In a Bondi coordinate and tetrad system one can find the Bondi
energy-momentum loss equations directly from the asymptotic Weyl tensor
equation\cite{Carlos.Ted,Gil.Ted}, Eq.(\ref{evolution}),
\begin{eqnarray}
\Psi ^{{\large \cdot }} &=&\sigma ^{{\large \cdot }}(\overline{\sigma })^{%
{\large \cdot }}+k\phi _{2}^{0}\overline{\phi }_{2}^{0},  \label{PSIDOT} \\
k &=&2G/c^{4}.  \notag
\end{eqnarray}
By substituting, from Eq.(\ref{dipole3}),
\begin{equation*}
\phi _{2}^{*0}=-\frac{2q}{c^{2}}[v^{i\prime }Y_{1i}^{-1}-2v^{ij\prime
}Y_{2ij}^{-1}],
\end{equation*}
and, from Eq.(\ref{sigma2}),

\begin{equation}
\sigma ^{{\large \cdot }}=\sqrt{2}24\frac{v^{ij}}{c}Y_{2ij}^{2},
\label{sigmadot}
\end{equation}
then using, on the left side, Eq.(\ref{PSI.M.P}),

\begin{equation}
\Psi (w,\zeta ,\overline{\zeta })=-\frac{2\sqrt{2}G}{c^{2}}M-\frac{6G}{c^{3}}%
P^{i}Y_{1i}^{0}+...,
\end{equation}
we have, (after Clebsch-Gordon expansions, from Eqs.(\ref{YY1}) and (\ref%
{YY3})), the mass (energy) loss theorem
\begin{equation}
M^{\prime }=\delta M_{0}^{\prime }=-\frac{288}{5G}cv^{ij}v^{ij}-\frac{2}{%
3c^{5}}q^{2}v^{i\,\prime }v^{i\,\prime }-\frac{32q^{2}}{5c^{5}}v^{ij\prime
}v^{ij\prime },  \label{M'}
\end{equation}
or
\begin{equation}
E^{\prime }=(Mc^{2})^{\prime }=-\frac{288}{5G}c^{3}v^{ij}v^{ij}-\frac{2q^{2}%
}{3c^{3}}v^{i\,\prime }v^{i\,\prime }-\frac{32q^{2}}{5c^{3}}v^{ij\prime
}v^{ij\prime },  \label{Edot}
\end{equation}
and the momentum loss equation

\begin{equation}
P^{i\,\prime }=\frac{8\sqrt{2}q^{2}}{5c^{4}}v^{j\prime }v^{ij\prime }.
\label{P'=0}
\end{equation}
The $c$ has been explicitly introduced.

There are several important things to notice about these relations:

\begin{itemize}
\item the energy loss due to the electromagnetic radiation is \textit{\
precisely }the electric dipole radiation expression, [with the dipole moment
being
\begin{equation}
d^{i}=q\xi ^{i},  \label{dipolemoment}
\end{equation}
allowing us to identify $\xi ^{i}$ as the center of charge as well as center
of mass] and electric quadrupole radiation, if we identify the electric
quadrupole moment, $D^{ij\,},$ with our $\xi ^{ij}$ from
\begin{equation*}
D^{ij\,\prime }=24\sqrt{2}qc\xi ^{ij},
\end{equation*}
so that\cite{L.L.}
\begin{equation*}
\frac{32q^{2}}{5c^{3}}v^{ij\prime }v^{ij\prime }=\frac{D^{ij\,\prime \prime
\prime }D^{ij\,\prime \prime \prime }}{180c^{5}}
\end{equation*}
the classical quadrupole electromagnetic energy loss.

\item By comparing our term, from Eq.(\ref{Edot}), for the gravitational
energy loss, namely
\begin{equation*}
\frac{288}{5G}c^{3}v^{ij}v^{ij},
\end{equation*}
with the standard quadrupole radiation formula\cite{L.L.}, i.e.,
\begin{equation}
\frac{G}{5c^{5}}\{Q^{ij\,\prime \prime \prime }\}\{Q^{ij\,\prime \prime
\prime }\},  \label{L&L}
\end{equation}
we find the physical meaning to the variables, $\xi ^{ij},$ namely modulo
numerical factors, it is the 2$^{nd}$ time derivative of the mass quadrupole
moment:
\begin{eqnarray}
\xi ^{ij} &=&\frac{G}{12\sqrt{2}c^{4}}Q^{ij\prime \prime },  \label{Qdotdot}
\\
v^{ij} &=&\frac{G}{12\sqrt{2}c^{4}}Q^{ij\prime \prime \prime },  \notag
\end{eqnarray}
\end{itemize}

{\bf Remark 15.} {\it Note that both the gravitational quadrupole
moment $Q^{ij}$ and the electric quadrupole moment $D^{ij}$ are
defined from the RT variable $\xi ^{ij}.$ }

The equations of motion for the world-line arise when $P^{i},$ from Eq.(\ref%
{P^i}), is used in the momentum loss equation. Writing Eq.(\ref{P^i}) as
\begin{equation}
P^{i}=Mv^{i}-\frac{2}{3c^{3}}q^{2}v^{i\,\prime }+MG+q^{2}H,  \label{P^i*}
\end{equation}
where $M_{0}$ has been replaced by $M$ and the $G$ and $H$ contain $l=2$
terms (which will be neglected in this discussion), $P^{i\,\prime }=\frac{8%
\sqrt{2}q^{2}}{5c^{4}}v^{j\prime }v^{ij\prime }$ becomes
\begin{equation}
Mv^{i\,\prime }=\frac{2}{3c^{3}}q^{2}v^{i\,\prime \prime }+\frac{8\sqrt{2}%
q^{2}}{5c^{4}}v^{j\prime }v^{ij\prime }-M^{\prime }v^{i}  \label{P'}
\end{equation}
Using the mass loss equation, Eq.(\ref{M'}), it becomes the equations of
motion for the world-line
\begin{equation}
Mv^{i\,\prime }=\frac{2}{3c^{3}}q^{2}v^{i\,\prime \prime }+\frac{8\sqrt{2}
q^{2}}{5c^{4}}v^{j\prime }v^{ij\prime }+v^{i}[\frac{2}{3c^{5}}
q^{2}v^{j\,\prime }v^{j\,\prime }+\frac{32q^{2}}{5c^{3}}v^{kj\prime
}v^{kj\prime }+\frac{288}{5G}cv^{jk}v^{jk}].  \label{Motion0}
\end{equation}

Many of the terms in this equation are quite familiar from classical
electrodynamics. If we ignore the mass loss terms (the three cubic terms) we
have the well-known radiation reaction equations\cite{L.L.} whose solutions
have the troubling run-away exponential behavior. By adding in the just
electromagnetic radiation term (the first and second cubic term) we again
have a classical electrodynamic result\cite{Thirring} which was derived in
the past by model building and which required mass renormalization. This
term has the correct sign and structure to possibly suppress the run-away
behavior - though whether or not it does so is not clear. The gravitational
radiation term (the third cubic term), arising from electromagnetic
quadrupole radiation, simply add to the likelihood that the run-away
behavior could be suppressed by these equations. At this point it is not
clear what effect the quadratic term, which couples the $l=1$ and $l=2$
harmonics, has on the motion.

There are two important comments to make concerning Eq.(\ref{Motion0}), one
positive, the other negative:

\begin{itemize}
\item The negative comment is that we have cheated a bit, in that we have
included a cubic term in Eq.(\ref{Motion0}) that comes from the mass loss,
while throughout this work we have excluded cubic terms. Our defense is that
it arises naturally from the mass loss and has such a clear physical meaning.

\item The positive remark is that we have not used any model building, as
has been used in the classical electrodynamic derivations, with the often
used mass renormalization. We have here simply taken the Einstein-Maxwell
equations and applied them to the class of non-twisting algebraically
special metrics and obtained the classical equations with the additional
gravitational correction term. All the numerical factors are identical to
those of the model dependent derivations. We have nothing that even
resembles mass renormalization. Our mass is the Bondi mass seen at infinity.
\end{itemize}

It has been known that there have been difficulties\cite{Piotr}, as we
mentioned earlier, in giving a definite answer to the question of the end
state of an RTMaxwell solution with a non-vanishing charge. It appears
likely that this difficulty is associated with the delicate balance between
the run-away behavior and its suppression via the three radiation terms. We
hope to be able to study the numerical evolution of Eq.(\ref{Motion0}) in
the near future. From a physical point of view we know that charged
particles do not self-accelerate so there is some reason to believe that the
same would be true for this special class of fields.

\section{Conclusion}

In this work we have returned to the study of the beautiful class of metrics
known as the Robinson-Trautman and Robinson-Trautman-Maxwell equations and
tried to give them physical content. Many attempts over the past forty or so
years have been made to solve them - with very few exact solutions - most
work being on approximations. Existence theorems for the pure vacuum case
were given and much approximate work was done on them but there appears to
be little progress on the Maxwell version. There have been several attempts%
\cite{Drey,Tafel}, via approximations, to relate the pure RT metrics to the
more general class of Bondi metrics where one has an understanding of
asymptotic energy and momentum as well as the multipole structure. However,
to our knowledge, little or nothing along these lines was done for the RTM
equations.

In this work we have first studied the RTM equations, in their natural NU/RT
coordinates, and reduced them, using a truncated spherical harmonic
expansion and approximation, to a series of ODE's. Afterwards we applied a
method, recently developed, to approximate the transformation to Bondi
coordinates and finally obtain the Bondi energy-momentum 4-vector and its
evolution. By studying these equations, we were able to give physical
meaning to many of the RTM variables. In addition we obtained directly from
the field equations, with no model building, the classical particle
equations of motion with the radiation reaction force and terms for the
possible suppression of the run-away behavior.

There are several different directions in which we intend to extend this
work:

\begin{itemize}
\item We would like to redo the calculations done here consistently to third
order in $v/c$. The obstacle is the complexity of the calculations. In the
present work often we had to do the calculation two or three times to get
agreement.

\item We plan to generalize this work to type II algebraically special
metrics that now contain shear-free null geodesic congruences with non%
\textit{-vanishing twist}. It appears virtually certain from earlier work
and from the charged Kerr metric, that the twist leads to intrinsic angular
momentum. The present equations of motion should, we believe, generalize to
motion for spinning particles.

\item We hope to generalize the second (or planned third) order calculations
given here for the RTM equations to the general case of asymptotically flat
Einstein-Maxwell metrics. The condition of being algebraically special would
be replaced by the tetrad condition of finding and using null geodesic
congruences that are asymptotically shear-free\cite%
{Ftprints,Carlos.Ted,Gil.Ted}. To do this appears to be a quite formidable
task.
\end{itemize}

\section{Acknowledgments}

This material is based upon work (partially) supported by the National
Science Foundation under Grant No. PHY-0244513. Any opinions, findings, and
conclusions or recommendations expressed in this material are those of the
authors and do not necessarily reflect the views of the National Science.
E.T.N. thanks the NSF for this support. G.S.O. acknowledges the financial
support from CONACyT through Grand No.44515-F, VIEP-BUAP through Grant
No.17/EXC/05 and Sistema Nacional de Investigadores (SNI-M\'{e}xico). C.K.
thanks CONICET and SECYTUNC for support.

\section{Appendix A}

For completeness, though we have not needed them, the Maxwell field up to
second order with $l=2$ terms is given by

\begin{eqnarray*}
\phi _{0}^{*0} &=&0, \\
\phi _{1}^{*0} &=&q[1+\frac{144}{5}v^{ij}v^{ij}+(\sqrt{2}v^{i}-\frac{72}{5}
v^{j}v^{ij})Y_{1i}^{0}-(2\sqrt{2}v^{ki}-\frac{1}{2}v^{k}v^{i}-\frac{144}{7}
v^{kl}v^{il})Y_{2ki}^{0}], \\
\phi _{2}^{*0} &=&-q[2v^{k\prime }-\frac{60\sqrt{2}}{5}(v^{j}v^{kj})^{\prime
}-\frac{24\sqrt{2}}{5}v^{i}v^{ki\prime }-i(v^{i\prime }v^{j}+\frac{96}{5}
v^{lj}v^{il\prime })\epsilon _{ijk}]Y_{1k}^{-1} \\
&&+q[4v^{jl\prime }-\frac{3}{\sqrt{2}}v^{j}v^{l\prime }-\frac{336\sqrt{2}}{7}
v^{js}v^{ls\prime }-4i(v^{jk}v^{i\prime }-\frac{1}{3}v^{i}v^{jk\prime
})\epsilon _{ikl}]Y_{2jl}^{-1},
\end{eqnarray*}

with each expression being a function of $\tau .$

\end{document}